\documentclass[%
 reprint,
 amsmath,amssymb,
 aps,superscriptaddress
]{revtex4-1}

\usepackage{graphicx,natbib}
\usepackage{dcolumn}
\usepackage{mathrsfs}
\usepackage{amsmath}
\usepackage{bbold}
\usepackage{bm}

\begin{document}
\title{An $\textrm{SU}(2)$ geometric phase induced by periodically driven Raman process in an ultracold dilute Bose gas}

\author{Zekai Chen}
\email[]{zchen57@ur.rochester.edu}
\author{Joseph D. Murphree}
\author{Nicholas P. Bigelow}
\affiliation{Department of Physics and Astronomy, University of Rochester, Rochester, New York 14627, USA}
\affiliation{Center for Coherence and Quantum Optics, University of Rochester, Rochester, New York 14627, USA}

\date{\today}

\begin{abstract}
	We propose a practical protocol to generate and observe a non-Abelian geometric phase using a periodically driven Raman process in the hyperfine ground state manifold of atoms in a dilute ultracold gas. Our analysis is based upon recent developments and application of Floquet theory to periodically driven quantum systems. The simulation results show the non-Abelian gauge transformation and the non-commuting property of the $\textrm{SU}(2)$ transformation operators. Based on these results, we propose a possible experimental implementation with an ultracold dilute Bose gas.
\end{abstract}

\maketitle

\section{Introduction}
The geometric phase is a phase factor acquired by a quantum system during adiabatic cyclic evolution. In 1984, M. V. Berry systematically discussed the geometric phase in non-degenerate quantum systems, and such a $\textrm{U(1)}$ Abelian geometric phase (Berry phase) appears as a phase factor on the non-degenerate states\cite{berry1984quantal}. F. Wilczek and A. Zee generalized Berry's work to degenerate quantum systems and showed that a non-Abelian geometric phase can be obtained\cite{PhysRevLett.52.2111}. Geometric phases have been studied in a broad range of physical systems and they connect to both fundamental and practical applications. In condensed matter physics, the geometric phase and the corresponding gauge structure in the Bloch band are closely related to the quantum Hall effect\cite{ye1999berry,bruno2004topological,haldane2004berry,zhang2005experimental}. In quantum computing, the non-Abelian geometric phase can be used to construct non-Abelian holonomic gates\cite{pachos1999non}, which are the foundation of robust holonomic quantum computing. There are many experimental realizations of non-Abelian geometric gates in different quantum systems, such as trapped ions\cite{duan2001geometric,leibfried2003experimental}, superconducting qubits\cite{abdumalikov2013experimental,zhang2015fast,yan2019experimental}, and nitrogen vacancy (NV) centers\cite{zu2014experimental}.

In the study of cold atoms in optical lattices, a geometric phase and the related Berry curvature of the Bloch band have been investigated\cite{atala2013direct,aidelsburger2015measuring,flaschner2016experimental} and are closely related to the study of the topology of the Bloch bands. In continuous quantum gases, the effects caused by the non-Abelian geometric phase have also been studied in a $^{87}\textrm{Rb}$ Bose-Einstein condensate (BEC), where the cyclic evolution of the atomic system was driven by slowly varying microwave and radio-frequency (RF) fields\cite{bharath2018singular,sugawa2018second}. Using a resonant tripod scheme, the non-Abelian adiabatic geometric transformation in the dark state manifold has also been realized in a metastable neon atom system\cite{theuer1999novel} and a cold strontium gas system\cite{leroux2018non}.

To obtain the non-Abelian geometric gauge transformation and non-Abelian gauge potentials, a quantum system with degenerate energy levels is necessary. The degenerate multi-level system in the study of continuous quantum gases is usually introduced by considering a multipod scheme\cite{ruseckas2005non,dalibard2011colloquium} or a system with a special symmetry property\cite{sugawa2018second,campbell2011realistic}. Recent theoretical works on the Floquet analysis of periodically driven systems shows that a periodically driven Hamiltonian can make quantum levels within the same Floquet band degenerate within the adiabatic approximation, and therefore one can realize non-Abelian geometric phase effects from a periodically driven system\cite{PhysRevA.95.023615,PhysRevA.100.012127}.

In this work we propose a practical experimental protocol for generating an $\textrm{SU}(2)$ non-Abelian geometric gauge transformation by a periodically driven Raman process and observing the $\textrm{SU}(2)$ geometric phase in a pseudo-spin-1/2 system in the ground state manifold of a non-interacting ultracold Bose gas system, where $\textrm{SU(2)}$ is the group of rotations of a spin-1/2 system and such a geometric phase results in a spin rotation in our system. Our analysis is based on the recent theoretical works\cite{PhysRevA.95.023615,PhysRevA.100.012127}, in which the authors applied Floquet theory to a system consisting of a spin interacting with a periodically driven magnetic field. They showed that when the oscillating magnetic field vector is slowly changing in direction, a non-Abelian geometric phase will appear. We build on this theoretical result by considering a pseudo-spin system interacting with a synthetic magnetic field generated by an optical Raman coupling, and propose a experimental protocol that may be realized practically. Our simulation shows that it is possible to measure the non-Abelian geometric phase using parameter sets that lie within the capabilities of existing cold atom experiments. Furthermore, we show that with our protocol one can observe the non-Abelian geometric phase even in the presence of undesired parameter fluctuations. 

Our pseudo-spin-1/2 system consists of two Zeeman sublevels in the hyperfine ground state manifold of an alkali atom. The periodically driven Raman process is realized by applying the product of a low frequency and a high frequency periodic signal simultaneously to the bias magnetic field, Raman laser intensities and relative phase between Raman lasers. Our simulation of the time-dependent Schr\"{o}dinger equation (TDSE) shows that an $\textrm{SU}(2)$ geometric phase can be observed and that the evolution operators which generate the geometric phase are non-Abelian, i.e. $[U_1,U_2]\ne0$, where $U_1$ and $U_2$ are different unitary transformation operators caused by different geometric phases. Although we use an $^{87}\textrm{Rb}$ system as an example, this protocol can be used in other atomic systems, both Bosonic and Fermionic, and has the potential to become a robust quantum control method.

\section{Non-Abelian geometric phase in a periodically driven system}
The Floquet analysis of periodically driven quantum systems has been well studied. We consider a system based on the periodically driven systems studied in two recent papers\cite{PhysRevA.95.023615,PhysRevA.100.012127}. Consider a spin-1/2 system whose Hamiltonian takes the form
\begin{equation}
H(t,\omega t+\theta)=\tilde{H}(t)f(\omega t+\theta), \nonumber
\end{equation}
where $\tilde{H}(t)=\tilde{H}[\vec{\lambda}(t)]$ is a Hamiltonian that depends on a set of slowly varying parameters $\vec{\lambda}(t)=\{\lambda_{\mu}\}$ ($\mu=1,2,3,...$), and $f(\omega t+\theta)$ is a periodic function with driving frequency $\omega$ (period $T=2\pi/\omega$) and phase offset $\theta$. For simplicity, we only consider the harmonic driving case, where $f(\omega t+\theta)=\cos(\omega t+\theta)$. The evolution of the state follows the Schr\"{o}dinger equation
\begin{equation}\label{Schrodinger}
i\hbar\partial_{t}|\psi(t)\rangle=H(t,\omega t+\theta)|\psi(t)\rangle.
\end{equation}
We can transform the system into a Floquet basis by introducing a micromotion operator $R(t,\omega t+\theta)=\exp\{iS(\omega t+\theta)\}$, where $S(\omega t+\theta)=\tilde{H}(t)\sin(\omega t+\theta)/\hbar\omega$ is the kick operator. The micromotion operator transforms the system from the physical basis into the Floquet basis.

Furthermore, we let $\tilde{H}(t)$ take the form
\begin{equation}\label{SU2 H form}
\tilde{H}=\hbar\Omega_0\hat{r}(t)\cdot\hat{\sigma},
\end{equation}
where $\hat{\sigma}=(\sigma_x,\sigma_y,\sigma_z)^{T}$ is the vector of Pauli operators, and $\hat{r}(t)=\hat{r}[\vec{\lambda}(t)]$ is a unit vector parameterized by $\vec{\lambda}(t)$. If we set $\Omega_{0}$ as constant, then the Hamiltonian $\tilde{H}$ describes a spin-1/2 system subject to a magnetic field whose direction is slowly changing. The Hamiltonian in the Floquet basis takes the form
\begin{eqnarray}
H_{F}&&=R^{\dagger}(t,\theta')H(t,\theta')R(t,\theta')-i\hbar R^{\dagger}(t,\theta')\partial_{t}R(t,\theta') \nonumber \\
&&=-\dot{\lambda}_{\mu}(i\hbar R^{\dagger}(t,\theta')\partial_{\mu}R(t,\theta')) \\
&&=\dot{\lambda}_{\mu}A_{\mu}(\vec{\lambda},\theta'), \nonumber
\end{eqnarray}
where we defined $\theta'=\omega t+\theta$, $\partial_{\mu}=\partial/\partial\lambda_{\mu}$ for fixed $\theta'$, and $A_{\mu}(\vec{\lambda},\theta')=-i\hbar R^{\dagger}(t,\theta')\partial_{\mu}R(t,\theta')$ is the gauge potential in the parameter space.

According to Floquet theory, the Hamiltonian in the Floquet basis can be written as a Fourier expansion of the form
\begin{equation}
H_{F}=\sum_{l}H_{F}^{(l)}e^{il\theta'}; l=0,\pm 1,\pm 2,..., \nonumber
\end{equation}
where $H_{F}^{(l)}=\frac{1}{2\pi}\int_{0}^{2\pi}H_{F}e^{-il\theta'}d\theta'$ is the $l$th Fourier component of $H_{F}$. If we assume the amplitudes of the matrix elements of these Fourier components are much smaller than the driving frequency, i.e.
\begin{equation}\label{adiabatic condition}
|\langle \alpha|H_{F}^{(l)}|\beta\rangle|\ll\hbar\omega,
\end{equation}
then this adiabatic condition allows us to neglect transitions between different Floquet bands and the evolution of the system can be approximated by the evolution within a single Floquet band\cite{PhysRevA.100.012127}. Furthermore, we can restrict our attention to the zeroth ($l=0$) Floquet band, since the states in the $l\ne0$ Floquet bands will evolve like their corresponding states in the $l=0$ band except for an additional $\textrm{U}(1)$ global phase factor $\phi_{global}=l\omega t$. Thus within the adiabatic approximation, the Hamiltonian in the Floquet basis is well approximated by the zeroth order Fourier component and can be written as
\begin{equation}\label{zeroth order effective Hamiltonian}
H_{F}^{(0)}=\dot{\lambda}_{\mu}A_{\mu}^{(0)},
\end{equation}
where $A_{\mu}^{(0)}$ is the zeroth order gauge potential $A_{\mu}^{(0)}(\vec{\lambda})=\frac{1}{2\pi}\int_{0}^{2\pi}A_{\mu}(\vec{\lambda},\theta')d\theta'$.

The zeroth order component of the Hamiltonian in the Floquet basis does not depend on the fast driving and thus can be regarded as the effective Hamiltonian of the spin-1/2 system in the Floquet basis. It takes the explicit form\cite{PhysRevA.100.012127}
\begin{eqnarray}\label{effective Hamiltonian}
H_{eff}&&=\frac{\hbar(1-J_0(a))}{2}\vec{r}\times\dot{\vec{r}}\cdot\vec{\sigma} \nonumber \\
&&=\dot{\lambda}_{\mu}\hbar g\epsilon_{ijk}r_i\partial_{\mu}r_j\sigma_k \\
&&=\dot{\lambda}_{\mu}A_{\mu}^{(0)},\nonumber
\end{eqnarray}
where $J_{0}(a)$ is the zeroth order Bessel function of the first kind, $a=|2\Omega_{0}(t)|/\omega$, $\epsilon_{ijk}$ is the Levi-Civita tensor (here $i$,$j$, and $k$ stand for $x$, $y$, and $z$), $g=\frac{1}{2}(J_{0}(a)-1)$, and we have used $\partial_{t}=\dot{\lambda}_{\mu}\partial_{\mu}$ with fixed $\theta'$. Note that the zeroth order $\textrm{SU}(2)$ gauge potential is $A_{\mu}^{(0)}=\hbar g\hat{n}\cdot\hat{\sigma}$, where $A_{\mu}^{(0)k}=\hbar g\epsilon_{ijk}r_i\partial_{\mu}r_j$ and $n_{k}=\epsilon_{ijk}r_{i}\partial_{\mu}r_{j}$.

The state in Floquet space is defined as $|\chi(t)\rangle=R^{\dagger}(t,\theta')|\psi(t)\rangle$, and the time evolution of the system in the Floquet space under adiabatic approximation can be written as
\begin{equation}
|\chi(t)\rangle=U_{eff}(t,t_0)|\chi(t_0)\rangle,
\end{equation}
or in the original basis
\begin{equation}
|\psi(t)\rangle=R(t,\theta')U_{eff}(t,t_0)R^{\dagger}(t_0,\theta'_0)|\psi(t_0)\rangle, \nonumber
\end{equation}
where $U_{eff}=\mathcal{T}\exp\{-\frac{i}{\hbar}\int_{t_0}^{t}H_{eff}dt'\}$ is the time evolution operator under the adiabatic approximation in the Floquet basis; and here $\mathcal{T}$ stands for time-ordering, and $\theta'_0=\omega t_0+\theta$.

Using Eqn.(\ref{effective Hamiltonian}), we can change the time evolution operator $U_{eff}$ from its time-ordered form into the form of a path-ordered $\textrm{SU}(2)$ unitary transformation operator $U_{eff}=\mathcal{P}exp\{-\frac{i}{\hbar}\int_{\vec{\lambda}(t_0)}^{\vec{\lambda}(t)}A_{\mu}^{(0)}d\lambda_{\mu}\}$, where $\mathcal{P}$ stands for path-ordering. The form of $U_{eff}$ shows that the evolution of the state only depends on the path that the parameter $\vec{\lambda}$ takes from its initial value $\vec{\lambda}(t_0)$ to its final value $\vec{\lambda}(t)$, and does not depend on its rate of change. If the parameter $\vec{\lambda}$ has cyclic time dependence, then the path-ordered unitary transformation operator only depends on the geometry of the closed loop that the parameter follows. In this case, the path-ordered operator takes the form $U_{c}=\mathcal{P}\exp\{-\frac{i}{\hbar}\oint A_{\mu}d\lambda_{\mu}\}$, and the system gains an $\textrm{SU}(2)$ geometric phase (Wilczek-Zee phase).

Unlike the $\textrm{U}(1)$ Berry phase that acts as a commutable phase factor, the non-Abelian geometric phase can cause population transfer between two eigenstates and two non-Abelian geometric phase factors related to different closed loops in the parameter space do not necessarily commute. Usually, to observe the adiabatic non-Abelian geometric phase, the system needs to be degenerate in order that all states acquire the same dynamic phase, which would otherwise make the geometric phase hard to detect. Therefore many studies of non-Abelian geometric phases are done within the degenerate dark state manifold. However, the physical system defined by $\tilde{H}$ in this work is not required to be degenerate. The periodic driving $f(\omega t+\theta)$ on the Hamiltonian $\tilde{H}$ introduces a Floquet band structure to the system, and the energy levels within the same Floquet band become degenerate under the adiabatic approximation\cite{PhysRevA.100.012127,PhysRevA.95.023615}. Since we set our parameters in the adiabatic regime, the system stays in the same Floquet band, and the cyclic evolution in the same Floquet band results in a non-Abelian geometric phase.

\section{Periodically driven Hamiltonian of the atomic system}
\begin{figure}
	\begin{center}
		\includegraphics[scale=0.4]{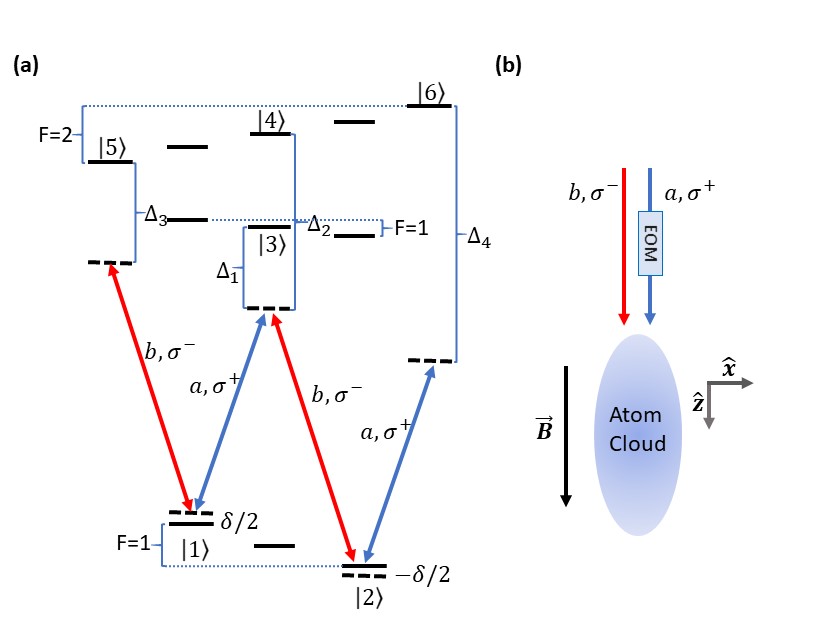}
	\end{center}
	\caption{(a) Level diagram of the Raman process that we consider. We choose $|1\rangle=|F=1,m_F=-1\rangle$ and $|2\rangle=|F=1,m_F=1\rangle$ in $5^{2}P_{\frac{1}{2}}, F=1$ manifold as our pseudo-spin-1/2 system. The ground states are coupled by two Raman lasers $a$ and $b$ with $\sigma^{+}$ and $\sigma^{-}$ polarizations, respectively. The excited states in the $5^{2}P_{\frac{1}{2}}, F=1$ and $F=2$ manifolds that couple to the ground states are $|3\rangle=|F=1,m_F=0\rangle$, $|4\rangle=|F=2,m_F=0\rangle$, $|5\rangle=|F=2,m_F=-2\rangle$ and $|6\rangle=|F=2,m_F=2\rangle$. $\Delta_{i}$, $i=1,2,3,4$ are single-photon detunings, $\delta$ is the two-photon detuning. (b) The experimental setup we consider: two Raman lasers merge to form a single beam before they interact with the atoms.}\label{level diagram}
\end{figure}
In this section we consider strategies for creating the periodic Hamiltonian discussed in the previous section using a Raman process in the ground state manifold of a $^{87}\textrm{Rb}$ atom. The Raman process has a variety of applications in the study of ultracold atoms, including quantum state manipulation, generating artificial gauge potentials and spin-orbit coupling, and creating topological defects\cite{wright2008raman,schultz2014raman,lin2011spin,goldman2014light,schultz2016creating,schultz2016raman}. We consider $|1\rangle=|F=1,m_F=-1\rangle$ and $|2\rangle=|F=1,m_F=1\rangle$ in the $5^{2}S_{\frac{1}{2}}, F=1$ ground state manifold of $^{87}\textrm{Rb}$ to be our pseudo-spin-1/2 system, as shown in Fig.\ref{level diagram}(a). The Raman process is realized by applying two co-propagating circularly polarized lasers to the ultracold atoms, which are subject to a weak bias magnetic field oriented along the beam axis ($z$-axis)\cite{wright2008raman,schultz2014raman,schultz2016creating,schultz2016raman}. 
Since the Raman lasers that we consider couple our pseudo-spin-1/2 states in the $5^{2}S_{\frac{1}{2}}, F=1$ manifold to the excited states in the $5^{2}P_{\frac{1}{2}}, F=1$ and $F=2$ hyperfine levels, the level diagram that we use in our calculation is actually a $W$-type instead of $\Lambda$-type\cite{wright2008raman}, see Fig.\ref{level diagram}(a). 

We start with the dipole interaction Hamiltonian $H_0=H_a-\vec{d}\cdot\vec{E}$, where $H_a$ is the Hamiltonian of the atom in the presence of a bias magnetic field, $\vec{d}$ is the atomic dipole moment, and $\vec{E}$ is the laser electric field, taking the form $\vec{E}=\vec{E}_ae^{-i\omega_a t}+\vec{E}_be^{-i\omega_b t}+c.c$, where $\omega_a$ and $\omega_b$ are laser frequencies. To solve the problem we can go to a rotating frame defined by the gauge transformation operator $U=\textrm{diag}\{e^{i\alpha_1},e^{i\alpha_2},e^{i\alpha_3},e^{i\alpha_4},e^{i\alpha_5},e^{i\alpha_6}\}$, where we define
\begin{align}
&\alpha_1=(\omega_1-\delta/2)t,&& \alpha_2=(\omega_2+\delta/2)t,  \nonumber \\
&\alpha_3=(\omega_3-\Delta_1)t,&& \alpha_4=(\omega_4-\Delta_2)t, \nonumber \\
&\alpha_5=(\omega_5-\Delta_3)t,&& \alpha_6=(\omega_6-\Delta_4)t; \nonumber
\end{align}
$\hbar\omega_i$ ($i=1,2,3,4,5,6$) is the energy of state $|i\rangle$ in the bias magnetic field; and we also define the one-photon detuning $\Delta_i$ ($i=1,2,3,4$) and two-photon detuning $\delta$ as
\begin{eqnarray}
&&2\pi\delta=\omega_a-\omega_b+\omega_1-\omega_2 \nonumber \\
&&2\pi\Delta_1=\omega_3-\frac{\omega_1+\omega_2}{2}-\frac{\omega_a+\omega_b}{2} \nonumber \\
&&2\pi\Delta_2=\omega_4-\frac{\omega_1+\omega_2}{2}-\frac{\omega_a+\omega_b}{2}  \\
&&2\pi\Delta_3=\omega_5-\frac{\omega_1+\omega_2}{2}-\frac{\omega_a+\omega_b}{2} \nonumber \\
&&2\pi\Delta_4=\omega_6-\frac{\omega_1+\omega_2}{2}-\frac{\omega_a+\omega_b}{2}. \nonumber
\end{eqnarray}
If we are in the far-detuned regime, namely where the one-photon detuning is much larger than the decay time of the excited state, we can adiabatically eliminate the excited states and get the effective two-level Hamiltonian\cite{wright2008raman,schultz2016raman,schultz2014raman,schultz2016creating}
\begin{equation}\label{W hamiltonian}
W=-\hbar\left( {\begin{array}{cc}
	\xi_{11}+\frac{\delta}{2} & \eta_{12}e^{-i\phi} \\
	\eta_{12}e^{i\phi} & \xi_{22}-\frac{\delta}{2} 
	\end{array} } \right),
\end{equation}
where the matrix elements are defined as
\begin{eqnarray}
\xi_{11}&&=\frac{|\Omega_{a13}|^2}{\Delta_1}+\frac{|\Omega_{a14}|^2}{\Delta_2}+\frac{|\Omega_{b15}|^2}{\Delta_3} \nonumber \\
\xi_{22}&&=\frac{|\Omega_{b23}|^2}{\Delta_1}+\frac{|\Omega_{b24}|^2}{\Delta_2}+\frac{|\Omega_{a26}|^2}{\Delta_4} \nonumber \\
\eta_{12}&&=\frac{|\Omega_{a13}\Omega_{b23}|}{\Delta_1}+\frac{|\Omega_{a14}\Omega_{b24}|}{\Delta_2} \nonumber.
\end{eqnarray}
Here $\Omega_{\rho ij}$ is the Rabi frequency, and it takes the form $\Omega_{\rho ij}=-d_{D1}E_{\rho}C_{ij}/\hbar$, with $\rho=a,b$, $i=1,2$, and $j=3,4,5,6$. $d_{D1}$ is the effective dipole moment of the $D_{1}$ transitions, $E_{\rho}$ is the electric field, $C_{ij}$ is the Clebsch-Gordon coefficient between states $|i\rangle$ and $|j\rangle$ and $\phi=\phi_{b}-\phi_{a}$ is the relative phase between the two Raman lasers. 

Our goal is to construct the periodically driven Hamiltonian with a high frequency driving signal $H=\tilde{H}f(\omega t+\theta)$, or in the harmonic driving case, $H=\tilde{H}\cos{\omega t}$, where $\theta$ is taken to be zero for simplicity. Notice that we can rewrite the effective two-level Hamiltonian Eqn.(\ref{W hamiltonian}) as
\begin{eqnarray}\label{SU(2) form of W}
W=&&-\frac{\hbar\left[\delta+(\xi_{11}-\xi_{22})\right]}{2}\sigma_z-\hbar\eta_{12}\cos{\phi}\sigma_x-\hbar\eta_{12}\sin{\phi}\sigma_y \nonumber \\
&&-\frac{\hbar(\xi_{11}+\xi_{22})}{2}\mathbb{1},
\end{eqnarray}
where $\mathbb{1}$ is the $2\times 2$ identity matrix. Since we are in the far-detuned regime, the single-photon detunings are much larger than the two-photon detuning, the Rabi frequencies, and the Zeeman splitting between different magnetic sublevels. If we set the Rabi frequencies to satisfy $|\Omega_{a13}|=|\Omega_{b23}|$, $|\Omega_{a14}|=|\Omega_{b24}|$, and $|\Omega_{a26}|=|\Omega_{b15}|$, with a bias magnetic field (e.g. 5G), then $\xi_{11}$ and $\xi_{22}$ will be approximately equal. Therefore we can further reduce the effective two-level Hamiltonian and write it as
\begin{equation}
W\approx-\hbar(\frac{\delta}{2}\sigma_z+\eta_{12}\cos{\phi}\sigma_x+\eta_{12}\sin{\phi}\sigma_y), \nonumber
\end{equation}
where we ignored the last term in Eqn.(\ref{SU(2) form of W}) since it only effects a $\textrm{U}(1)$ global phase factor during the evolution. 

To create the harmonic driving of the Hamiltonian, we can modulate the bias magnetic field and $\eta_{12}$ as $\cos{\omega t}$. In this case, the effective two-level Hamiltonian $W$ can be regarded as the desired driven Hamiltonian $H$, and takes the form
\begin{equation}
H=-\hbar(\frac{\tilde{\delta}}{2}\sigma_z+\tilde{\eta}_{12}\cos{\phi}\sigma_x+\tilde{\eta}_{12}\sin{\phi}\sigma_y)\cos{\omega t}, \nonumber
\end{equation}
where $\tilde{\delta}\cos{\omega t}=\delta$ and $\tilde{\eta}_{12}\cos{\omega t}=\eta_{12}$. Now let $\tilde{\delta}=2\Omega_0\cos{\Theta(t)}$ and $\tilde{\eta}_{12}=\Omega_0\sin{\Theta(t)}$, where $\Theta(t)$ is the slowly varying parameter. This can be achieved by driving both Raman lasers as $\sqrt{|\sin{\Theta(t)}\cos{\omega t}|}$ and changing the relative phase $\phi=\phi_b-\phi_a$ between two lasers from $\phi=\Phi$ to $\phi=\pi+\Phi$, where $\Phi$ is a parameter that does not depend on fast periodic driving. The periodic Hamiltonian can finally be written in the desired form
\begin{equation}\label{effective periodic Hamiltonian}
H(t)=\hbar\Omega_0\hat{r}\cdot\hat{\sigma}\cos{\omega t},
\end{equation}
with $\hat{r}(\vec{\lambda})=(-\sin{\Theta}\cos{\Phi},-\sin{\Theta}\sin{\Phi},-\cos{\Theta})^{T}$. If we fix $\Phi$ and slowly drive $\Theta$ in a cyclic manner, namely $\Theta=\Omega t$, where $\Omega\ll\omega$, we will obtain the desired Hamiltonian (Eqn.(\ref{SU2 H form})) that leads to an $\textrm{SU}(2)$ non-Abelian geometric phase. Note that we take $\vec{\lambda}=\{\Theta,\Phi\}$ as the set of coordinates on a unit 2-sphere with a time dependent polar angle $\Theta=\Omega t$ and a fixed azimuthal angle $\Phi$. This Hamiltonian describes a pseudo-spin-1/2 system in a rotating synthetic magnetic field whose magnitude is modulated.

\section{A Practical experimental protocol and simulation Results}
\subsection{The geometric phase}
After realizing the effective two-level periodic Hamiltonian in Eqn.(\ref{effective periodic Hamiltonian}), the dynamics of the system in the rotating frame follow from the Schr\"{o}dinger equation (Eqn.(\ref{Schrodinger})). Using Eqn.(\ref{effective Hamiltonian}), we find the zeroth Fourier component of the effective Hamiltonian in the Floquet basis to be
\begin{equation}
H_{eff}=\hbar\Omega g\hat{n}\cdot\hat{\sigma},
\end{equation}
where $\hat{n}=(-\sin{\Phi},\cos{\Phi},0)^T$, and $g=\frac{1}{2}(J_{0}(a)-1)$ with $a=2\Omega_0/\omega$. We assume $\Theta=\Omega t$, $\Phi=\textrm{const}.$ so that $\dot{\Theta}=\Omega$, $\dot{\Phi}=0$. Therefore, the zeroth order $\textrm{SU}(2)$ gauge potential takes the form
\begin{equation}
A_{\Theta}^{(0)}=-\hbar g(\sin{\Phi}\sigma_x-\cos{\Phi}\sigma_y).
\end{equation}
We can write the $\textrm{SU}(2)$ transformation operator as
\begin{equation}\label{evolution operator}
U(t,t_0)=\exp\{ig(\sin{\Phi}\sigma_x-\cos{\Phi}\sigma_y)\left[\Theta(t)-\Theta(t_0)\right]\}.
\end{equation}
For cyclic evolution, we get the geometric phase $\gamma=2m\pi g$, $m=\pm1,\pm2,\pm3,...$, and the $\textrm{SU}(2)$ transformation operator can be written as
\begin{equation}\label{cyclic transformation operator}
U_c=\exp\{-i\gamma\vec{n}\cdot\vec{\sigma}\}=\exp\{i2m\pi g(\sin{\Phi}\sigma_x-\cos{\Phi}\sigma_y)\},
\end{equation}
where $m$ is an integer. In the case that we consider, $m$ takes a negative sign due to the choice of states.

We use a fourth-order finite-difference method to solve the time-dependent Schr\"{o}dinger equation (TDSE) (Eqn.(\ref{Schrodinger})). The TDSE (Eqn.(\ref{Schrodinger})) describes the dynamics in the rotating frame, but the evolution operator that causes the geometric phase is in the Floquet basis. Notice that the micromotion operator that can transform the rotating basis to the Floquet basis takes the form $R=\exp\{i\tilde{H}(t)\sin{\omega t}/\hbar\omega\}$, and it goes to the identity operator at the end of each cycle of fast driving, namely $\sin{\omega T_q}=0$, with $T_q=2q\pi/\omega$ ($q=0,1,2,3,...$). Thus if we prepare the system into an eigenstate of $\tilde{H}(t_0)$ at the initial time $t_0$ and turn on the Raman laser and periodic driving abruptly, the system will start evolving in the Floquet basis. If we measure the system at the end of each fast driving cycle, the rotating basis will be already aligned with the Floquet basis so we get a direct measurement in Floquet space.

\subsection{Experimental setup}

\begin{figure}
	\begin{center}
		\includegraphics[scale=0.3]{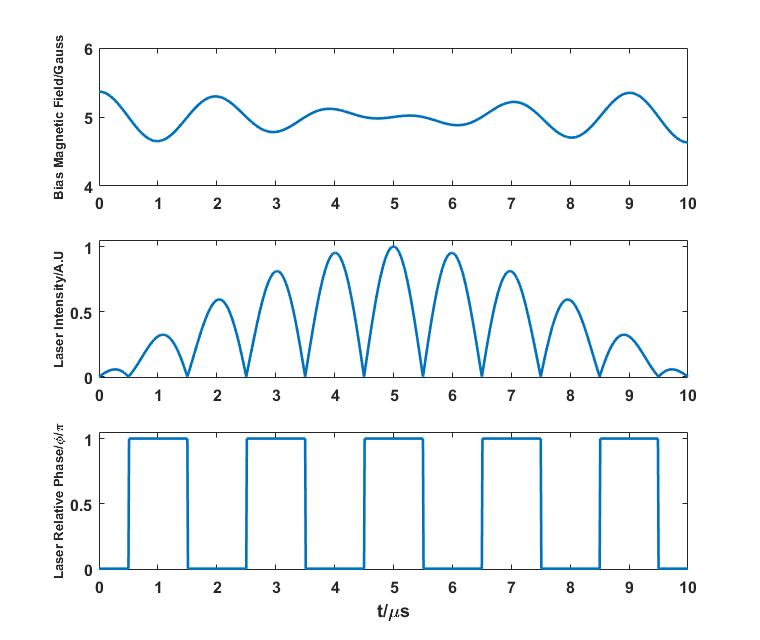}
	\end{center}
	\caption{From top to bottom: (a) driving signal of the magnetic field, (b) laser intensity, and (c) relative phase between lasers. The driving parameters are $\Omega=2\pi\times50\textrm{kHz}$, $\omega=2\pi\times500\textrm{kHz}$. Here we only show the driving signals for the first $10\mu s$.}\label{driving profiles}
\end{figure}

To experimentally realize the setup we consider, two co-propagating Raman lasers along the z-axis with left and right circular polarizations are needed, see Fig.\ref{level diagram}(b). Unlike the usual Raman process, where the Raman laser intensities are time-independent or only have slow time dependence, here we need to modulate both the laser intensities and the relative phase between two Raman lasers with a periodic function that is the product of a low frequency and a high frequency periodic signal. Meanwhile the magnitude of the bias magnetic field generated by a Helmholtz coil also needs to be modulated to provide the periodic two-photon detuning $\delta$. The driving signal of the bias magnetic field is $B=B_0+\Delta B\cos{\Omega t}\cos{\omega t}$ and the Raman laser intensities are proportional to $|\sin{\Omega t}\cos{\omega t}|$, respectively, see Fig.\ref{driving profiles}.(a,b). The relative phase between two Raman lasers follows the function $\frac{\pi}{2}[1-\textrm{sgn}(\sin{\Omega t}\cos{\omega t})]$, where $\textrm{sgn}(x)$ is the signum function, as shown in Fig.\ref{driving profiles}.(c).

To realize the desired periodically driven signal of the parameters, we can first use an arbitrary waveform generator (AWG) to generate the modulation signals. Then we can propagate the signals to acousto-optical modulators (AOMs) to drive the intensity of the Raman lasers and to an electro-optical modulator (EOM) to drive the relative phase between the Raman lasers. The parameters that we choose in our simulation are $\Omega=2\pi\times50\textrm{kHz}$ and $\omega=2\pi\times500\textrm{kHz}$ with the beam waist $w=300\mu m$, which will not push commercial AOMs beyond their limits. Finally, the modulation of the bias magnetic field can be realized by sending the modulation signal from the AWG to an audio power amplifier, and use it to drive the Helmholtz coil that generates the bias magnetic field.

\subsection{Preparation of the states and projection measurements}
After describing a potential experimental setup, we discuss the preparation of the initial states and how to do projection measurements of the desired quantum state. Our theoretical framework in this paper is based on a single atom, which assumes the ultracold Bose gas is dilute enough so that the interaction between atoms is negligible. In this work, we consider an ultracold dilute $^{87}\textrm{Rb}$ Bose gas and we focus on the $5^{2}S_{\frac{1}{2}}, F=1$ ground state manifold. 

We produce a Bose-Einstein condensate in the $|1\rangle\equiv|R=1,m_F=-1\rangle$ state and then must transfer the population to the desired initial state. There are many ways to control the system and achieve this state preparation. In our laboratory, we have developed a reliable Raman waveplate method to achieve state rotations on the Bloch sphere, and we can use a Raman waveplate pulse to rotate the states and measure the atomic Stokes parameters\cite{schultz2016creating,hansen2016singular,schultz2014raman}. The waveplate pulse couples states $|1\rangle$ and $|2\rangle$ and rotates the system to the initial state $|\psi(0)\rangle$, which is a superposition state of $|1\rangle$ and $|2\rangle$. Other than the Raman waveplate, there are also other ways to transfer the atoms into the desired initial state, such as using a radio-frequency pulse sequence.

We can use a Stern-Gerlach time-of-flight (TOF) imaging method to measure the populations in different states. To get phase information from the system, we need to rotate the system into the eigenbasis of the $x$ and $y$ axes. This can be achieved by any high-fidelity $\pi/2$ rotation operation around $x$ and $y$ axes, i.e. Raman waveplate pulses. Generally, if we ignore the undetectable global phase, the state of the system can be written as $|\psi\rangle=c_1|1\rangle+c_2e^{i\beta}|2\rangle$, where coefficients $c_1$ and $c_2$ are real, and $\beta$ is the relative phase. The atomic Stokes parameters are defined as
\begin{eqnarray}
&&S_1=2c_1c_2\cos{\beta} \nonumber\\
&&S_2=2c_1c_2\sin{\beta} \\
&&S_3=c_1^2-c_2^2, \nonumber
\end{eqnarray}
and they can be understood as projection measurements on the $x$, $y$ and $z$-axes of the Bloch sphere. The Stern-Gerlach TOF can be regarded as a measurement of $S_3$, while to measure the other two atomic Stokes parameters $S_1$ and $S_2$ we need to apply a $\frac{\pi}{2}$-waveplate pulse to rotate the detection axis to the $x$ and $y$-axes\cite{hansen2016singular}. From the atomic Stokes parameters we are able to extract both the population and phase information of the state.

The Stern-Gerlach TOF imaging happens after we turn the Raman lasers off, so we are performing the measurement in a Zeeman basis defined solely by the bias magnetic field. However, in our calculations we work in a rotating frame. Thus our final state, which is stationary in the rotating frame, will acquire an extra phase factor between the $|1\rangle$ and $|2\rangle$ states of $\alpha=\alpha_2-\alpha_1=(\omega_a-\omega_b)t$ in the Zeeman basis. Since in our experimental setup, the laser frequencies are fixed and shifted by AOMs, we are able to record the frequency difference between Raman lasers. Therefore, we can calculate the extra phase difference at any time when the Raman lasers are on and eliminate the extra phase factor in data processing.

\subsection{Simulation results}
\begin{figure}
	\begin{center}
		\includegraphics[scale=0.35]{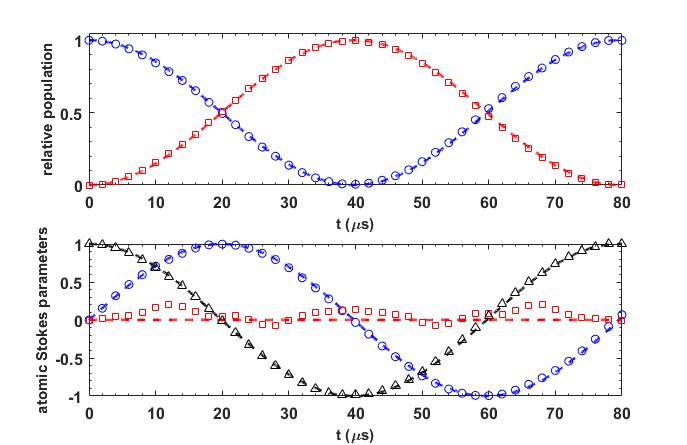}
	\end{center}
	\caption{Simulation results of an $\textrm{SU}(2)$ transformation, where circles and squares and triangles are simulation results from solving the TDSE in the rotating frame, and dashed lines are analytical curves calculated in the Floquet basis under adiabatic approximation. Upper plot: population transfer, blue and red represent $|1\rangle$ and $|2\rangle$, respectively.  Lower plot: evolution of the atomic Stokes parameters, where blue, red and black represent $S_1$, $S_2$, and $S_3$, respectively. Here we use the parameters: maximum laser powers $P_a=P_b=271.7\mu W$, beam waist $w=300\mu m$, magnitude of the time-varying part of the bias magnetic field $\Delta B=0.368G$, bias magnetic field average $B_0=5G$, which results in $\Omega_0=2\pi\times258.3\textrm{kHz}$. The driving frequencies are $\Omega=2\pi\times50\textrm{kHz}$ and $\omega=2\pi\times500\textrm{kHz}$. With the above parameters, the geometric phase result in a Rabi-like oscillation with a period $T_{geo}\approx80\mu s$. }\label{geometric phase}
\end{figure}
By numerically solving the TDSE Eqn.(\ref{Schrodinger}), we get the evolution of the system in the rotating frame by extracting the points at the end of each fast driving cycle. Also, we analytically calculate the evolution of the state subjected to the $\textrm{SU}(2)$ unitary transformation given by Eqn.(\ref{evolution operator}) in the Floquet basis under the adiabatic approximation. As shown in Fig.\ref{geometric phase}, the simulation results match the analytical calculations well. In the upper plot, blue circles and red squares are the simulation results states of $|1\rangle$ and $|2\rangle$, respectively. The initial state is $|\psi(0)\rangle=|1\rangle$. Dashed lines are results of the analytical calculation for population transfer. In the lower plot, blue circles, red squares and black triangles are simulation results for atomic Stokes parameters $S_1$, $S_2$, and $S_3$, respectively. Dashed lines are the analytical predictions. The parameters we use are: maximum laser powers $P_a=P_b=271.7\mu W$, beam waist $w=300\mu m$, magnitude of the time-varying part of the bias magnetic field $\Delta B=0.368G$, the bias magnetic field average $B_0=5G$, which results in $\Omega_0=2\pi\times258.3\textrm{kHz}$. The driving frequencies are $\Omega=2\pi\times50\textrm{kHz}$ and $\omega=2\pi\times500\textrm{kHz}$. With the above parameters, $g=-0.1248$, and the geometric phase produces a Rabi-like oscillation with a period $T_{geo}\approx80\mu s$.

Using the atomic Stokes parameter values at the end of the slow driving cycle, we can calculate the $\textrm{SU}(2)$ transformation operator $U_c$. Take $t=20\mu s$ as an example. The atomic Stokes parameters take values $S_1=0.998$, $S_2=0.06315$, and $S_3=-0.0014$, which give $|c_1|^2=0.4993$, $|c_2|^2=0.5007$, and $\beta=0.063\textrm{rad}$. Therefore, the $\textrm{SU}(2)$ transformation operator at $t=20\mu s$ is calculated to be $U_c=0.7066\mathbb{1}-i(0.0445\sigma_x+0.7062\sigma_y)$, which matches the analytical prediction $U_{c}^{theory}=(\mathbb{1}-i\sigma_y)/\sqrt{2}$ that describes a $\pi/2$ rotation around the $y$-axis on the Bloch sphere.

\begin{figure}
	\begin{center}
		\includegraphics[scale=0.35]{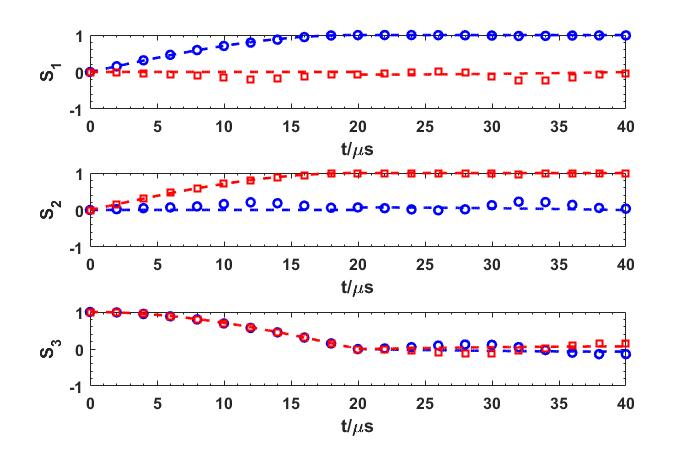}
	\end{center}
	\caption{Non-Abelian property of the evolution operators. From top to bottom: evolution of the atomic Stokes parameters $S_1$, $S_2$ and $S_3$, respectively. The evolution operators $U_1$ and $U_2$ are achieved by using the same parameters as Fig.(\ref{geometric phase}) and setting the relative phase between Raman lasers as $\Phi_1=0$ and $\Phi_2=\pi/2$, respectively. The duration of each evolution operator is $\tau=20\mu s$. Different colors represent the results for different order of operators, blue: $U_2U_1$ and red: $U_1U_2$. Blue circles and red squares represent simulation results from TDSE while dashed lines are theoretical predictions. The different final results for atomic Stokes parameters shows $[U_1,U_2]\ne0$, which proves the geometric phase that we get is non-Abelian.} \label{non-Abelian}
\end{figure}

After solving for the the $\textrm{SU}(2)$ transformation operator, the next step is to prove the non-Abelian property of the geometric gauge transformations. We consider two $\textrm{SU}(2)$ transformation operators 
\begin{eqnarray}
U_1&&=\exp\{i2\pi g\sigma_y\}\approx\frac{1}{\sqrt{2}}(\mathbb{1}-i\sigma_y), \nonumber\\
U_2&&=\exp\{-i2\pi g\sigma_x\}\approx\frac{1}{\sqrt{2}}(\mathbb{1}+i\sigma_x), \nonumber
\end{eqnarray}
which are constructed by turning on the periodically driven Hamiltonian for $t=20\mu s$ and setting the relative phase parameter to be $\Phi_1=0$ and $\Phi_2=\pi/2$, respectively. All the other parameters are the same as what we used in Fig.\ref{geometric phase}. After constructing the $\textrm{SU}(2)$ transformation operators, we apply them to the initial state $|\psi(0)\rangle=|1\rangle$, one after another in different orders $U_2U_1$ and $U_1U_2$. As shown in Fig.\ref{non-Abelian}, the difference in the atomic Stokes parameters at the final time shows that the $\textrm{SU}(2)$ transformation operators that we construct do not commute, $[U_1,U_2]\ne0$, which verifies the non-Abelian property of the geometric phase.

The geometric phase that we get over one low frequency cycle depends on the parameters we choose, as long as the adiabatic condition Eqn.(\ref{adiabatic condition}) is satisfied. Therefore, we can easily change the parameters, such as Raman laser intensities, modulation amplitude of the magnetic field, and the periods of both low and high frequency driving parameters to tune the geometric phase over a broad range of values. In both simulation results, we see slight differences of the TDSE solution from the analytical predictions. This is the joint effect of the non-zero $\frac{1}{2}(\xi_{11}-\xi_{22})$ term in Eqn.(\ref{SU(2) form of W}) and the quadratic Zeeman effect, which was ignored in the construction of the effective two-level Hamiltonian of the Raman process. The non-zero $\xi_{11}-\xi_{22}$ term will introduce an additional term proportional to $|\sin{\Theta t}\cos{\omega t}|\sigma_z$ in the effective two-level Hamiltonian. The quadratic Zeeman shift will introduce an additional term to the two-photon detuning that is proportional to $(\cos{\Omega t}\cos{\omega t})^2$, which brings in additional zeroth and second harmonic terms. The zeroth harmonic from both terms can be canceled by shifting the laser frequencies, but the higher harmonic terms will bring in additional terms in the effective Hamiltonian and affect the dynamics of the system. However, such effects do not affect our result very much because we work with a weak bias magnetic field, and under the conditions that we consider in our simulations the amplitude of the additional higher harmonic terms is $2\pi\times1.88\textrm{kHz}$, whereas $\Omega_0=2\pi\times258.3\textrm{kHz}$. Therefore, the amplitudes of both $\frac{1}{2}(\xi_{11}-\xi_{22})$ term and the quadratic Zeeman shift will be much smaller than the amplitude of two-photon detuning $\delta$ so that both terms are negligible.

\subsection{Robustness against parameter fluctuations}
\begin{figure}
	\begin{center}
		\includegraphics[scale=0.2]{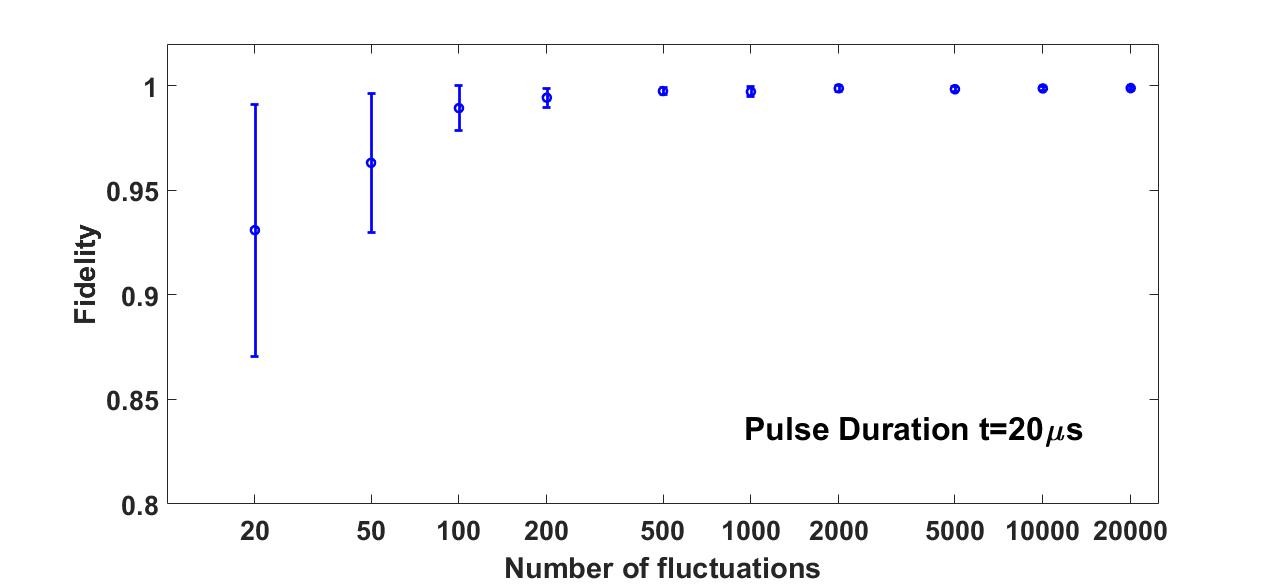}
	\end{center}
	\caption{Fidelity of the geometric gauge transformation versus number of fluctuations. Each point is calculated by averaging five runs with the same number of fluctuations for 26 different initial states, the errorbars is the standard deviation of all the fidelities calculated from the 130 runs for the same number of fluctuations. The ideal $\textrm{SU}(2)$ transformation operator is $U_{c}^{ideal}=(\mathbb{1}-i\sigma_y)/\sqrt{2}$. The standard deviation of random Gaussian noises for bias magnetic field and laser powers are $5\%$ of their amplitudes. For relative phase, the standard deviation of the Gaussian noise is $0.01\pi$. All fluctuations are distributed evenly during the pulse duration.} \label{noise}
\end{figure}
After showing the non-Abelian property of the geometric phase, it is natural to ask if such a geometric phase is sufficiently robust against parameter fluctuations that one can observe it in the laboratory. We introduce random Gaussian noise to the magnitude of the bias magnetic field, the Raman laser powers, and the relative phase between the two Raman lasers. The ideal $\textrm{SU}(2)$ transformation operator takes the form $U_{c}^{ideal}=(\mathbb{1}-i\sigma_y)/\sqrt{2}$ and causes a $\pi/2$ rotation around $y$ axis. With the parameters we considered in Fig.\ref{geometric phase}, the pulse duration is $20\mu s$. We denote the operator with noise as $U_c^{noise}$. The standard deviation of the magnetic field noise and laser power noise are $5\%$ of their amplitudes, while for the relative phase $\phi$, the standard deviation of the Gaussian noise is $0.01\pi$. Then we start from the same initial state $|\psi_0\rangle$, vary the number of fluctuations that are evenly distributed over $20\mu s$ and calculate the fidelity $f\equiv|\langle \psi_{ideal}|\psi_{noise}\rangle|^2$, where $|\psi_{noise}\rangle=U_c^{noise}|\psi_0\rangle$ and $|\psi_{ideal}\rangle=U_{c}^{ideal}|\psi_0\rangle$. The results are shown in Fig.\ref{noise}, where each point is the average fidelity calculated from 26 different initial states evenly distributed on the Bloch sphere with 5 runs for each state, and the error bars are the standard deviations calculated from the $5\times26=130$ data sets for each number of fluctuations. We can see that the average fidelity is always above $0.9$, which shows that the operators we constructed with noise are robust against random fluctuations for the different initial states that we consider. However, we see much smaller error bars for the numbers of fluctuations higher than $200$, which correspond to high frequency fluctuations (above $10\textrm{MHz}$), than we see for lower frequency fluctuations (below $10\textrm{MHz}$). Since for low frequency fluctuations, $5\%$ parameter fluctuation can greatly deform the contour that the rotating synthetic magnetic field takes and therefore have a greater influence on the resulting geometric phase. For higher frequency noise, the averaged fidelity is above $98\%$ and the error bars become much smaller than the low frequency points, which indicates that the deformation of the contour in parameter space is averaged out and the geometric phase becomes robust against high frequency random fluctuations.

The fluctuations we consider here are all above $1\textrm{MHz}$. Since our pulse duration is relatively short, the lower frequency fluctuations of up to several kilohertz, such as mechanical noises, can be regarded as long term drift with small amplitude for our problem, which does not degrade the observation of our effect significantly. The high fidelity shown in Fig.\ref{noise} demonstrates that the non-Abelian geometric phase induced by the periodically driven Raman process is robust enough to be observed and therefore has the potential to be another method to control the quantum state of a cold atom system.

\section{Conclusion and discussion}
In this paper we have proposed a possible realization of a periodically driven Hamiltonian through periodically driving a Raman process in the hyperfine ground state manifold of an alkali atom. A non-Abelian geometric phase is observed according to our simulation results. Through measuring the atomic Stokes parameters, we are able to measure the $\textrm{SU}(2)$ transformation operator of the cyclic evolution in Floquet space. The non-commuting property of two different $\textrm{SU}(2)$ transformation operators subject to different geometric phase factors proves the non-Abelian property of this geometric phase. For simplicity, we only set one of our parameters $\Theta$ as time dependent. In fact the other parameter that we consider $\Phi$ can also be time dependent as long as the parameters form a closed loop. Based on the general theory of a spin interacting with a periodically driven magnetic field\cite{PhysRevA.100.012127,PhysRevA.95.023615}, our work extends the realization of the non-Abelian geometric phase into a pseudo-spin system with Raman coupling. We also proposed a practical experimental implementation using a dilute ultracold atomic gas interacting with Raman lasers and we verify that the non-Abelian geometric phase effect can be observed in the laboratory even with the presence of possible parameter fluctuations.

Due to the oscillating magnetic field, an additional quadratic Zeeman effect term will appear in the Hamiltonian\cite{gan2018oscillating}. However, as we discussed, the effect caused by the quadratic Zeeman effect can be ignored when the field is sufficiently weak. Another issue is heating. Since we use a Raman process with large single-photon detuning, the dynamics of the system is confined to the ground state manifold, so the heating caused by spontaneous emission from the excited states is negligible. The excitation between different Floquet bands is also suppressed if we let the parameters satisfy the adiabatic condition Eqn.(\ref{adiabatic condition}). In addition, in our protocol we use co-propagating Raman lasers, so there will be no momentum transfer in our periodically driven Raman process. Therefore the Floquet heating effect discussed in \cite{li2019floquet} will be suppressed as well. Finally, in our analysis, since the duration of the evolution is much less than the typical decoherence time of ultracold Bose gases, we can ignore decoherence effects and use pure state descriptions of the system in the calculations. For a more general case, if the duration of the geometric phase pulse becomes comparable to the decoherence time, one needs to take decoherence into consideration and use density matrix methods in instead.

Our simulation results show that by introducing periodically driven interactions, one can promote the Abelian physical system to a non-Abelian one under the adiabatic approximation\cite{PhysRevA.100.012127,PhysRevA.95.023615}, and that the associated geometric phase is sufficiently robust against parameter fluctuations and thus detectable in the laboratory. To obtain the non-Abelian geometric phase and the non-Abelian gauge potential, one needs to work with a system with degenerate quantum levels. As discussed by Novi{\v{c}}enko\cite{PhysRevA.100.012127}, the non-Abelian geometric phase comes from neglecting the transitions between different Floquet bands such that the states in the same Floquet band become degenerate. There are other strategies to get non-Abelian geometric phase effects\cite{leroux2018non,sugawa2018second} that work in the degenerate eigenbasis of a system. Such an eigenbasis consists of superposition states in the Zeeman basis. In contrast, the experimental protocol that we propose realizes a non-Abelian geometric phase effect in a Floquet basis that can be projected to the Zeeman basis, allowing us to measure the system in a more direct way.  Although we only considered a $^{87}\textrm{Rb}$ Bose gas in this work, the protocol we propose can be used in other atomic or ionic systems, both Bosonic and Fermionic. Therefore, our protocol has the potential to be a reliable quantum control method in ultracold atom studies.

\section{Acknowledgement}
We thank Maitreyi Jayaseelan and Elisha Haber for discussions. We also thank Gediminas Juzeli\ifmmode \bar{u}\else \={u}\fi{}nas for the useful comments on the manuscript. This work was supported by the National Science Foundation grant number PHY-1708008 and NASA/JPL RSA 1616833.

\bibliography{reference}
\end{document}